\documentclass{aa}  
\usepackage{color}
\usepackage{ulem}
\usepackage{soul}
\usepackage{sidecap}
\usepackage{comment}
\AddToHook{begindocument/before}{\usepackage{hyperref}}

\usepackage{graphicx}
\usepackage{txfonts}
\begin{document}

   \title{Onset of penumbra formation}


   \author{ M. Garc\'ia-Rivas\inst{1,2},
          J. Jur\v c\'ak\inst{1},  
          N. Bello Gonz\'alez\inst{3},
          J.M. Borrero\inst{3}, 
          R. Schlichenmaier\inst{3},
          \and
          P. Lindner\inst{3}   
          }

   \institute{  Astronomical Institute of the Czech Academy of Sciences, Fri\v cova 298, 25165 Ond\v rejov, Czech Republic \\
                \email{marta.garcia.rivas@asu.cas.cz}    
                                \and
                Astronomical Institute, Charles University, V Hole\v sovick\'ach 2, 18000 Prague, Czech Republic
                \and
                Institut f\"ur Sonnenphysik (KIS), Sch\"oneckstrasse 6, 79104 Freiburg, Germany
             }

   \date{Received 11 28, 2023; accepted 03 08, 2024}

 
  \abstract
   {Fully fledged penumbrae have been widely studied both observationally and theoretically. Yet the relatively fast process of penumbra formation has not been studied closely with high spatial resolution.}
   {We investigate the stages previous to and during the formation of penumbral filaments in a developing sunspot.}
   {We analysed Milne-Eddington inversions from spectro-polarimetric data of the leading sunspot of NOAA\,11024 during the development of its penumbra. We focused on selected areas of this protospot in which segments of penumbra develop.}
   {We find that few types of distinctive flow patterns develop at the protospot limb and centre sides previous to penumbra formation. 
   The flow in the centre side is often characterised by a persistent (>20\,min) inflow-outflow pattern extending radially over 4\,arcsec at the direct periphery of the protospot umbra. This inflow-outflow system often correlates with elongated granules, as seen in continuum intensity maps, and is also coupled with magnetic bipolar patches at its edges, as seen in magnetograms. The field is close to horizontal between the bipolar patches, which is indicative of its possible loop configuration. All of these aspects are analogous to observations of magnetic flux emergence. In the protospot limb side, however, we observed a mostly regular pattern associated with small granules located near the protospot intensity boundary. Locally, an inflow develops adjacent to an existing penumbral segment, and this inflow is correlated with a single bright penumbral filament that is brighter than filaments containing the Evershed flow. All investigated areas at the centre and limb side eventually develop penumbral filaments with an actual Evershed flow that starts at the umbral boundary and grows outwards radially as the penumbral filaments become longer in time.

   }
    {}
   
   \authorrunning\space
   \maketitle
   \nolinenumbers
%

\section{Introduction}

As discussed by \cite{BelloGlez+2019}, the basic necessary conditions of a minimum required flux and the presence of strong horizontal fields for the formation of penumbrae are known in general. However, there are a number of aspects that are still not known and/or understood. Observational analyses of pores and sunspots imply that a minimum magnetic flux on the order of 10$^{20}$~Mx is necessary for the penumbra to form \citep{Zwaan1987, Leka1998}. This number is consistent with the theoretical model by \citet{Rucklidge1995}, who investigated the stability of magnetic flux tubes and their properties based on their total flux. These models and observations describe pores and sunspots created by rather symmetrical magnetic flux tubes. As the magnetic flux within the flux tube increases, the magnetic field inclination on its edges increases, and the horizontal component of the magnetic field is strong enough to shape the rising convective cells to form the elongated penumbral filaments. The magnetohydrodynamic (MHD) simulations of sunspots by \citet{Rempel2009} as well as analysis of the observations by \citet{Jurcak+etal2014} suggest that the critical magnetic field inclination necessary for penumbra formation is around 45$^\circ$. This value is also in agreement with the simpler models of \citet{Rucklidge1995}.

Another property of penumbrae in sunspots that might play a role in sustaining horizontal fields resulting in penumbra formation is the presence of an overlaying magnetic field that traps the emerging horizontal field in the photosphere in the form of penumbral filaments. \citet{Solanki:1993} proposed the uncombed structure of the magnetic field in sunspot penumbra where horizontal flux tubes or tops of elongated convective cells (as suggested by MHD simulations) are interleaved by a stronger and more vertical magnetic field known as the background magnetic field or intraspines. The actual configuration of the magnetic field is very similar to the proposed model, as confirmed by \citet{Tiwari_etal2013}, and it was shown by \citet{borrero2007,Borrero:2008} that the background magnetic field closes above the penumbral filaments. 

Another form of penumbra is the orphan penumbra. These are penumbrae not coupled to an umbra. \cite{Jurcak+etal2014} reported no background magnetic field present in these structures, while \citet{Zuccarello+etal2014} found indications of it. Orphan penumbrae are in most aspects (e.g. Evershed flows and penumbral grains) similar to sunspot penumbrae \citep{Jurcak+etal2014}. \citet{Lim+etal2013} and \citet{Zuccarello+etal2014} found orphan penumbrae originating from an emerging magnetic field trapped at the photospheric level by the pre-existing overlying fields in areas of complex topology in active regions, typically near the polarity inversion lines.

Details of penumbra formation were studied by \citet{Schlichenmaier+etal2010a, Schlichenmaier+etal2010b, Schlichenmaier+etal2011, Schlichenmaier+etal2012}, who used unique observations of sunspot formation over a period of 4.5~h. The formation of penumbra was in this case caused by the accumulation of magnetic flux on one side of the protospot where a stable segment of a penumbra formed on the opposite side. Prior to the formation of a stable penumbra, \citet{Schlichenmaier+etal2011} observed temporal filaments with what they introduced as counter-Evershed flows. Using the same data, \citet{Rezaei+etal2012} found signatures of a low lying (observed in the Fe\,I 617.3\,nm line) magnetic canopy in the regions where penumbra later formed. This scenario was subsequently confirmed by the numerical simulations of \citet{MacTaggart+etal2016}, who studied the flux emergence of a twisted magnetic flux tube and the magnetic canopy formed preferentially away from the emergence site. Indication of a chromospheric magnetic canopy was found also by \citet{Shimizu+etal2012} and \citet{Romano+etal2013}, who observed a chromospheric halo in regions where penumbra later formed. The presence of a chromospheric halo preceding the formation of the penumbra was interpreted by \citet{Romano+etal2013, Romano+etal2014}, and \citet{Murabito+etal2016} as a necessary configuration of the magnetic field that allows for downward bending of the chromospheric field lines and the creation of a shallow loop that is also associated with temporal counter-Evershed flows caused by siphon flows. The recent work by \cite{Lindner_2023} favours a scenario in which emerging flux blocked by an overlying (chromospheric) canopy provides the strong horizontal fields necessary for penumbra formation.

\cite{Li+2018} investigated the formation of a penumbra sector and found that in a first stage, the area and magnetic flux of the forming penumbral sector and the umbra both increased and that in a subsequent second stage, the area and magnetic flux of the penumbra increased at the expense of the umbra area and the flux. They also found a persistent blueshift in the penumbra formation area that they interpreted as a signature of magnetic flux emergence. However, they did not find any chromospheric counterpart coupled to the penumbra formation unlike the annular region reported by \cite{Shimizu+etal2012}. Yet, they found a correlation between the penumbral sector formation and sunspot rotation. \cite{Li+2019} also found indications of penumbral areas formed by emerging flux trapped in the photosphere by the overlying pre-existing field in a complex active region.

\citet{Jurcak2011} discovered the magnetic property defining the umbra-penumbra boundary of stable sunspots, a critical value of the vertical magnetic field \citep[1.86\,kG,][]{Jurcak_etal2018}. \citet{Jurcak+etal2015} found this canonical value to be a natural inner boundary for the expansion of penumbral magneto-convection in umbral areas. Their result is in agreement with the findings of \cite{Li+2018}, as umbral areas are originally occupied by developing penumbra. Therefore, penumbra can only set in and operate in regions with a vertical field below the critical value. The observation of such a vertical magnetic field boundary against vigorous modes of magneto-convection (e.g. the penumbral mode) is supported by the analysis of the Gough and Tayler criterion \citep{Gough:1966} of stability by vertical fields against overturning convection by \cite{Schmassmann+2021}. 

From a modelling point of view, there are a number of realistic simulations of sunspots with penumbra \citep[See, e.g.,][]{Rempel2009, Rempel2011, Rempel2012, Chen_2017, Panja+2021}. Yet only a few works address the onset and formation of penumbra. In accordance with observations, \citet{Hurlburt+Rucklidge2000}, \citet{Simon+Weiss1970}, and \citet{Jahn+Schmidt1994} found that an increasing inclination (with respect to the normal to the solar surface) of the magnetic field and an increasing flux in sunspot models are key factors for the origin of penumbra. \citet{Wentzel1992} proposed the `fallen magnetic flux tubes' model for penumbra formation. There, the upwelling of a mass flow in the inner footpoint of the field lines within the umbra provokes the flux tubes to fall onto the photosphere and become submerged in the surroundings of the sunspot (i.e. the penumbra forms from top to bottom). On local scales, \cite{Rempel2011} presented a more elaborated model where the flux tubes bend due to the mass loads originated from pressure-driven upflows (see Fig. 19 of the cited work). On the other hand, \cite{Chen_2017} simulated the formation of an asymmetric sunspot formed by the coalescence of magnetic features with the same polarity from small-scale dipoles embedded in granular cells (i.e. the penumbra forms from bottom to top). Regarding penumbral radial flows (e.g. Evershed flow or counter-Evershed flow), \cite{Chen_2017} found that penumbral filaments were dominated by inflows in the region where the flux emergence was more active (between the two main spots). More interestingly, some penumbral filaments developed intermittent outflows even when they were originally dominated and surrounded by inflows. Transient inflows embedded in outflows were obtained by \cite{2018_Siu-Tapia} from simulations based on the MURaM radiative MHD code.

In this work, we describe the observational properties of magneto-convection at the outskirts of a protospot before the onset of a penumbra and during the first stages of the penumbral formation process. The observations and data analysis are described in Sec. \ref{sec:analysis}. Three distinct penumbral formation processes are respectively presented and discussed in Sec. \ref{sec:results} and Sec. \ref{sec_conclusions}. This work closes with conclusions obtained from the results.

\begin{figure*}
\centering
   \includegraphics[width=\linewidth]{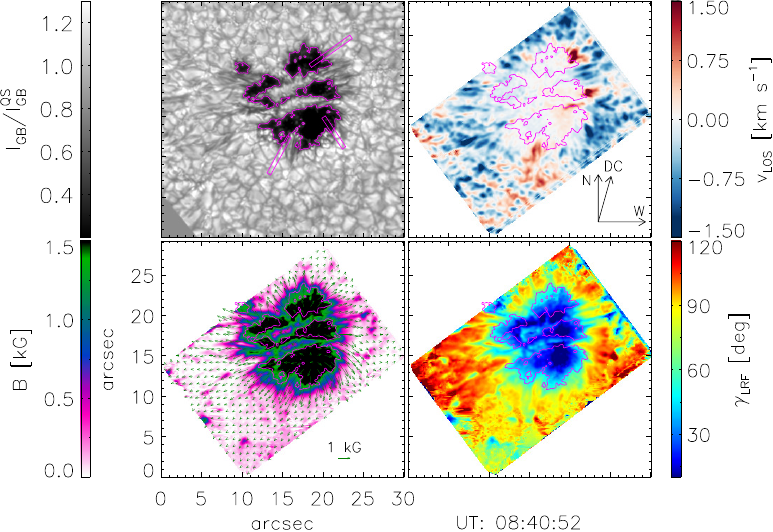}
\caption{Snapshot of NOOA\,11024 sunspot at the early stages of penumbra development (08:40:52\,UT) as seen in G-band intensity (upper left), LOS velocity (upper right), total magnetic field strength (lower left), and magnetic field inclination in the LRF (lower right). The pink iso-contour outlines the umbral boundary using a threshold of 50\% of the quiet-Sun intensity. The three pink boxes in the G-band intensity maps mark the regions studied in detail. The maps have been rotated according to the solar north-south direction. The temporal evolution is available \href{https://www.aanda.org/}{\textbf{online}}.}
\label{fullfov}
\end{figure*}

 \begin{figure}[!b]
   \centering
   \includegraphics[width=\linewidth]{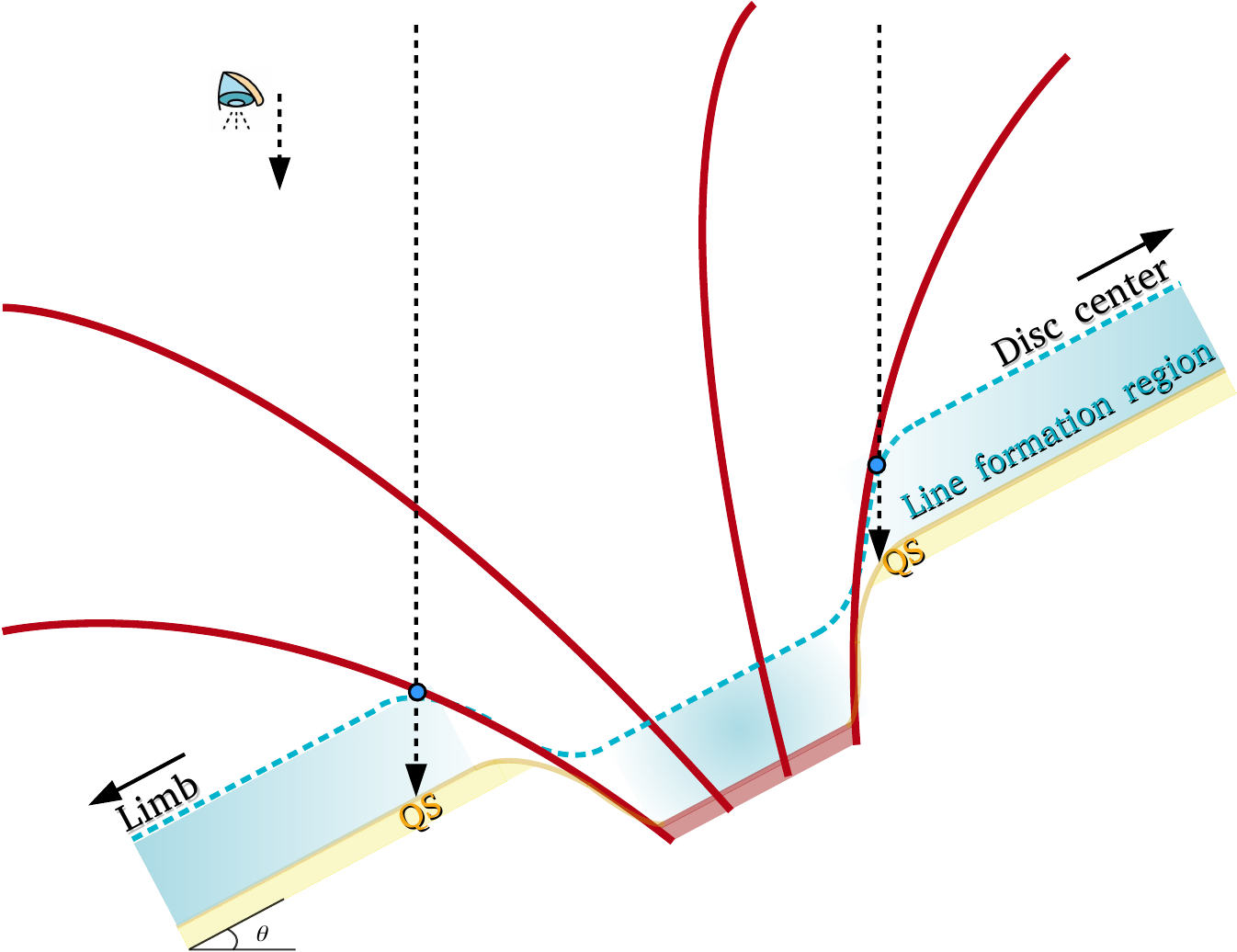}
     \caption{Sketch of the projection effects on the derived magnetic properties around a spot at $\theta = 28^\circ$. The magnetopause (i.e. the most external red lines) influences the derived magnetic properties of regions farther away from the umbral boundary at the limb side than at the centre side.  
     }
\label{fig:sketch_canopy}
\end{figure}

\section{Observations and data analysis} \label{sec:analysis}

The data were recorded on 9 July 2009 with the `G\"ottingen' Fabry-Pérot Interferometer 
\citep[GFPI,][]{bendlin92, puschmann06} in the full-Stokes polarimetric mode \citep{bello08} at the VTT Telescope (Tenerife). The observations consisted of a time series of 4:40\,h following the evolution of the leading sunspot in NOAA\,11024 located at a heliocentric angle $\mathrm{\theta \sim 28^\circ}$. Here, we make use of the imaging data in the G-band and full-Stokes polarimetry in the \ion{Fe}{i} 617.3\,nm photospheric line. We note that these data were previously used in other investigations \citep{Schlichenmaier+etal2010a, Schlichenmaier+etal2010b, Rezaei+etal2012, BelloGlez+2012, Jurcak+etal2014b}, and therefore we defer to the referenced works for further details about the observations.

We restricted our analysis to a 2\,h time span (08:32\, UT-10:32\,UT) of uninterrupted observation of penumbra formation consisting of 109 GFPI scans. We focused on selected regions of the protospot periphery in which the formation of penumbral segments can be followed over time. We ran Milne-Eddington inversions of the \ion{Fe}{i} 617.3\,nm spectropolarimetric observations with the VFISV code \citep{Borrero+etal2011}, which provides the components of the magnetic field vector and line-of-sight (LOS) velocity maps. We used the code AMBIG \citep{disambiguation_leka} to solve the 180$^{\circ}$ ambiguity in the component of the magnetic field perpendicular to the observer's LOS. We also made use of routines of the AZAM code \citep{Lites_etal1995} to transform the magnetic field vector from the LOS frame to the local reference frame (LRF). During the time series, the field of view varies. This resulted in a limited effective field of view.

   \begin{figure*}
\centering
 \sidecaption
   \includegraphics[width=0.9\linewidth]{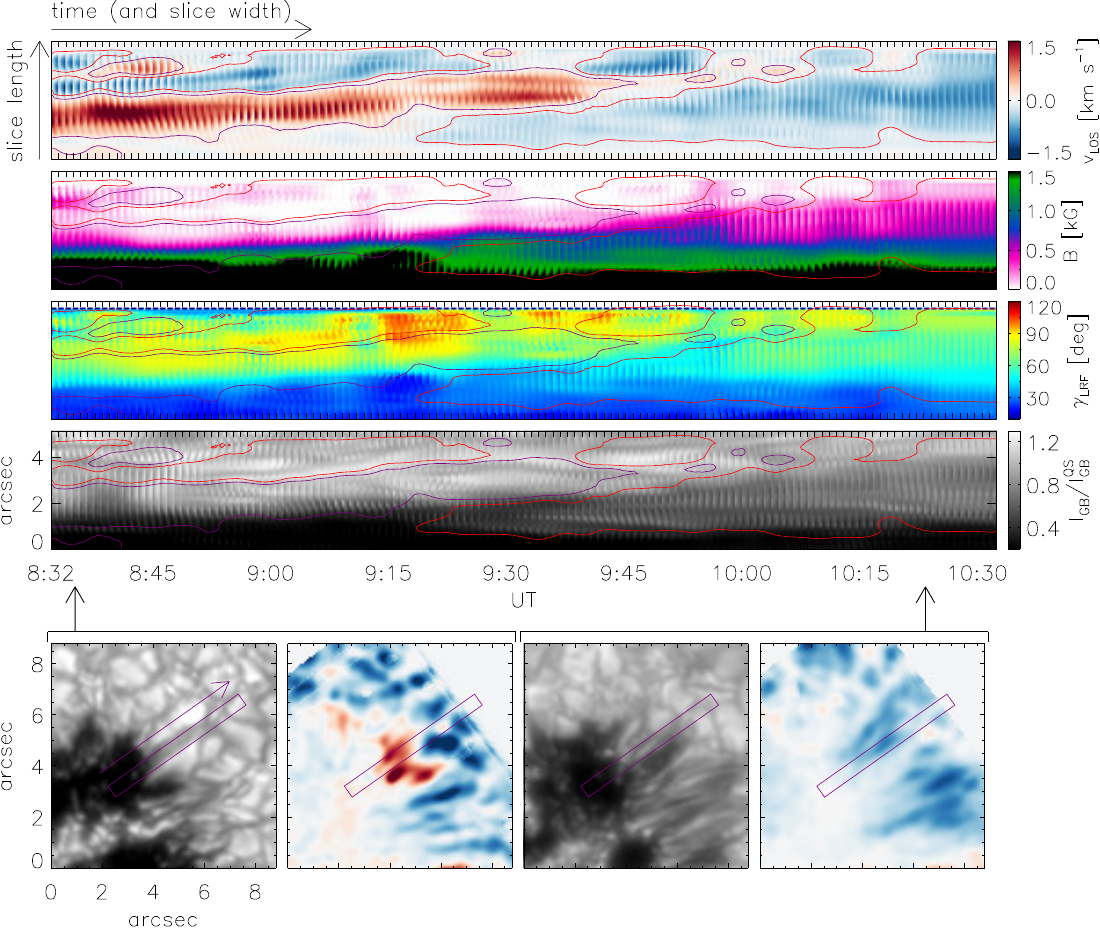}
 \caption{Bi-directional flow in elongated granules. Top panels: Time slices of selected penumbra-formation region on the sunspot centre side. From top to bottom: LOS~velocities, magnetic field strength, inclination (LRF), and G-band intensity. Bottom panels: Context maps containing the areas (marked by purple boxes) selected for the time slices as seen in the G-band intensity and LOS~velocity at 8:35\,UT (two left panels) and at 10:24\,UT (two right panels). The violet arrow points the direction along the slice length. The purple and red iso-contours highlight velocities of $\pm$130\, m\,s$^{-1}$, respectively.}
\label{timeslice_cs1}
\end{figure*}

\section{Results} \label{sec:results}

Figure\,\ref{fullfov} shows a snapshot of the field of view under analysis at the beginning of the observations. An animation of the development and evolution of the protospot can be found \href{https://www.aanda.org/}{\textbf{online}}. At this early stage, the protospot already shows segments of stable penumbra around (22\arcsec, 20\arcsec) and (16\arcsec, 11\arcsec). The penumbra around (11\arcsec, 19\arcsec) is highly distorted by the continuous emergence of the new flux in the developing active region \citep{Rezaei+etal2012}. We note that the protospot is the leading polarity of NOAA\,11024, and the following polarity is directed towards the east (not visible within the field of view).

We studied the magnetic field and velocity properties in regions where penumbrae later form. We note that the Milne-Eddington inversion scheme does not allow for disentangling of the different physical properties of the spot background component from the penumbral filaments. Instead, the Milne-Eddington inversion provides mean values of LOS velocity, magnetic field strength, and inclination \citep{borrero2014} from regions where the 617.3\,nm line forms and is most sensitive to changes of the physical properties. This implies that the magnetic canopy affects the plasma properties inferred at the external boundary of the umbral core and beyond. In this case, the influence is asymmetric due to the heliocentric angle and the resulting projection effects. As seen in the bottom maps of Fig. \ref{fullfov} ($B$ and $\gamma_{LRF}$), the magnetic boundary oriented towards the disc centre is sharper than the magnetic boundary oriented towards the solar limb (i.e. the magnetic radial gradient at the umbral boundary is larger at the centre side than at the limb side). Figure\,\ref{fig:sketch_canopy} illustrates how the projection of the magnetopause on the quiet-Sun regions induces a larger magnetic gradient at the centre side of the umbral boundary, while it induces a smaller magnetic gradient at the limb side, which translates into a sharper magnetic boundary at the centre side of the spot. We note that this is a simplified sketch where the complex nature of the magnetopause is not shown and the line formation region, spot size, and Wilson depression are not to scale.

In the following three subsections, we identify three distinct types of magneto-convection: a bi-directional flow in elongated granules located at the centre side of the spot, and a transient filament with counter-Evershed flow and a granular pattern located at the limb side. All of them are followed by the formation of the penumbra and the onset of the Evershed flow.

\subsection{Bi-directional flow in elongated granules}
\label{bidirectional}

On the centre-side region of the protospot, the penumbra starts to form at approximately 9:20\,UT (i.e. some 50\,min after the start of our observations). In Fig. \ref{timeslice_cs1}, we show a 2\,h time slice of a 5\,pixel-wide selected region from 08:32\,UT until 10:30\,UT for intensity, inclination, field strength, and LOS~velocity. 

The time slice shows that from the start of the observations and for about 50\,min, the velocity in the selected region at the very outskirts of the umbra is characterised by a persistent redshift-blueshift pattern (i.e. the flow is directed towards the umbra in its close vicinity and outwards further out). This bi-directional velocity pattern occurs in regions with weak magnetic fields of less than 500 G. Only the innermost end of the redshifted region is rooted in regions with 1000~G field strength inside the umbra, as seen in the continuum intensity slices. The time slice shows that before the penumbra sets in, the (radial) red-blue pattern is associated with inclinations that go from positive polarity (purple) to horizontal (white) to opposite polarity (green). The G-band intensity images show that the bi-directional flow pattern is associated with one intensity structure that resembles an elongated granule. The mean intensity of this granule is comparable to the quiet-Sun intensity. 

We note that in the region under study, the LOS inclination of the magnetic field is around 30$^\circ$ and towards the centre of the solar disc. The strong magnetic field observed close to the umbral boundary is part of the magnetic body of the spot. Further away from the umbral boundary, the LOS does not cross the magnetic canopy of the protospot, as illustrated in Fig. \ref{fig:sketch_canopy} (centre side), and thus we observe the actual magnetic field configuration of the bi-directional flow pattern.

From about 9:10\,UT onward, the red-blue patch is displaced away from the umbra, and the displacement increases in time. The reason for this displacement is the formation of a new convective cell featuring an outwards-oriented plasma motion (blueshifted region) that is initially most visible in the G-band continuum images as faint, bright filaments that do not have a strong enough LOS flow to be encircled by the red contour (i.e. $\mathrm{-130~m\,s^{-1}}$). This outwards flow can be considered the first appearance of the Evershed flow. As time progresses, the length of the region occupied by the incipient Evershed flow expands. In the continuum intensity slices, one can see multiple bright structures whose inner footpoint, which is also the brightest, moves inwards into the umbra similarly to penumbral grains in stable penumbrae \citep[e.g.][]{Sobotka:1999b}. From the intensity time slices it becomes obvious that the forming penumbra with Evershed flow does not heat the solar surface as effectively as the convection pattern characterised by the bi-directional flow. The mean G-band intensity in the forming penumbra decreases to 70\,\% of the quiet-Sun intensity (0.7\,$I_\textrm{QS}$). This value is comparable to the average penumbral intensity measured in fully developed sunspot penumbrae. 

Along with the expansion of the region occupied by the Evershed flow, we observed an expansion of the region with a strong magnetic field. The initially sharp gradient of the magnetic field strength at the protospot boundary became more diffuse. At the end of the analysed period, the magnetic field strength increased to approximately 600~G at the outer edge of the slice. A similar trend was also observed for the magnetic field inclination and can be explained by the expansion of the magnetic field that is associated with the penumbra formation. In the beginning, we had a sharp boundary between the nearly vertical magnetic field at the protospot boundary and the horizontal field in the weakly magnetised bi-directional flow patch. Later on, we found LRF inclinations around 80$^\circ$ at the end of the slice, and such values are comparable to those found at the outer penumbral boundaries of stable sunspots. 

\subsection{Transient filament with counter-Evershed flow}
\label{transient}

\begin{figure*}
\centering
\sidecaption
   \includegraphics[width=0.9\linewidth]{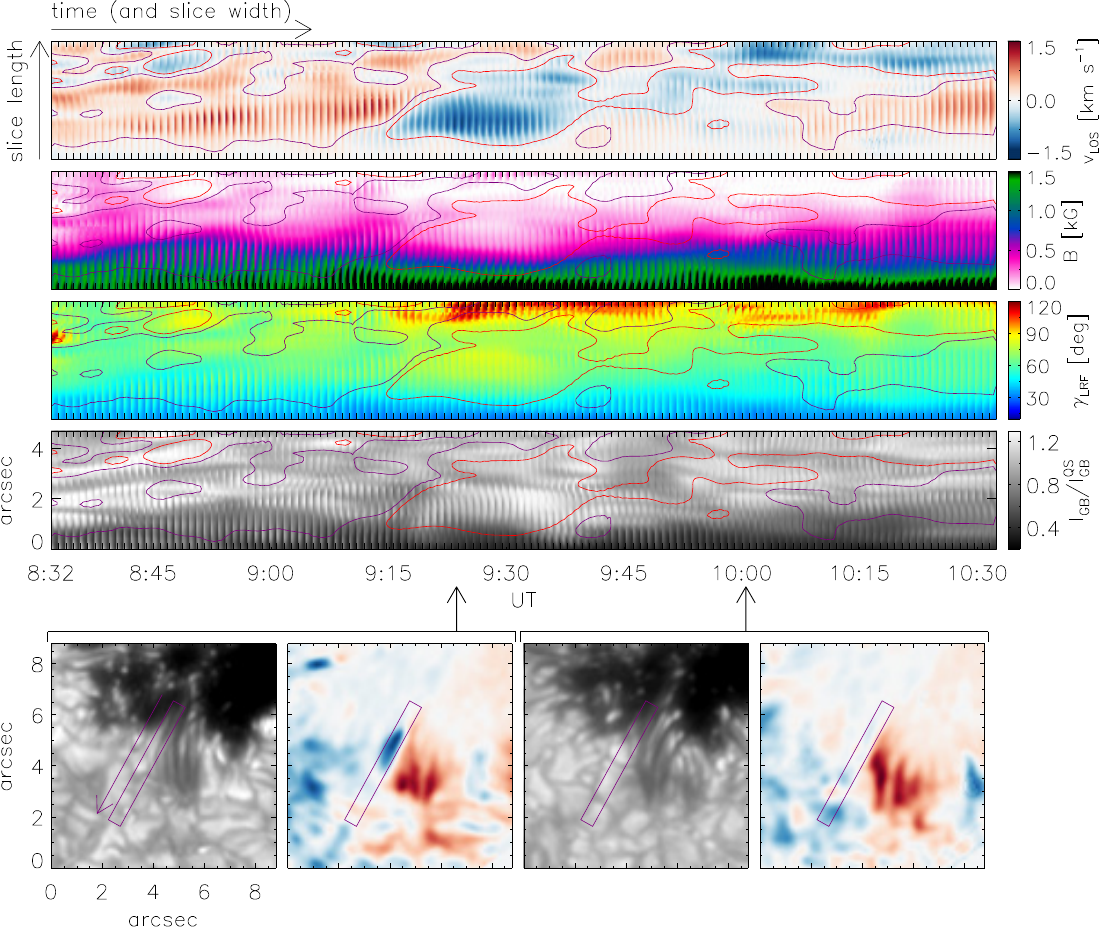}
 \caption{Analogous to Fig.~\ref{timeslice_cs1} but showing the limb-side region where a transient filament with a counter-Evershed flow formed. The bottom panels correspond to observations at 9:23\,UT (two left panels) and at 10:01\,UT (two right panels).}
\label{timeslice1}
\end{figure*}

On the limb-side boundary of the protospot, we observed one clear example of the formation of a transient filament carrying a counter-Evershed flow. In Fig.~\ref{timeslice1}, we show the time slice evolution of this region. We note that the LOS inclination in this region is around 70$^\circ$ and points towards the limb. Therefore, the magnetic boundary of the protospot is not as sharp as on the centre side, and our results are influenced by looking through the magnetic canopy of the protospot. Also, the viewing angle is such that upflows located at the penumbral bright grains are seen as blueshifts, and a regular Evershed flow is therefore not represented only by redshifts.

Prior to the formation of the transient filament, we observed within the slice an edge of the region with a formed penumbra. The innermost part of the penumbra had intensities around 0.7\,$I_\textrm{QS}$, but the second row of penumbral bright grains (located around 2\arcsec of the slice length) reached intensities around $I_\textrm{QS}$. As mentioned before, at this location, the bright grains are associated with blueshifts; these blue patches are visible until 8:52\,UT. The magnetic boundary of the protospot was diffuse due to the presence of the penumbra, while the magnetic field strength and inclination steadily decreased and increased along the slice, respectively.

Around 9:18\,UT, a filament with counter-Evershed flow developed. Within 10~minutes, it reached its full length and maximum LOS velocity and then decayed for approximately 15~minutes. It reached its maximum continuum intensity in later stages, around 9:34\,UT. Its formation can also be identified in the time-slice plots of the magnetic field strength and inclination. The magnetic field is weaker and more horizontal than that of regular penumbra observed before and afterward. A similar behaviour has been reported for some of the cases in a statistical study of counter-Evershed flows by \cite{Castellanos2021}. 
After the disappearance of the transient filament, we observed a similar magnetic field configuration as before its formation. There are no strong LOS velocities observed, as the LOS inclination is close to 90$^\circ$ within the slice. We note that the Evershed flow observed before 9:12\,UT is associated with LRF inclinations around 60$^\circ$, and such inclinations are not found within the slice between minutes 9:42\,UT and 10:02\,UT. Starting at 9:52\,UT, we observed weak blueshifts associated with bright penumbral grains of very short penumbral filaments that have LOS inclinations close to 90$^\circ$. After 10:02\,UT, the filaments became longer and more inclined, so the redshifted regions associated with Evershed flow appeared. As in the case of \ref{bidirectional}, we observed steady lengthening of the filaments carrying the Evershed flow in time.

\subsection{Granular pattern} 

\begin{figure*}
\centering
\sidecaption
   \includegraphics[width=0.9\linewidth]{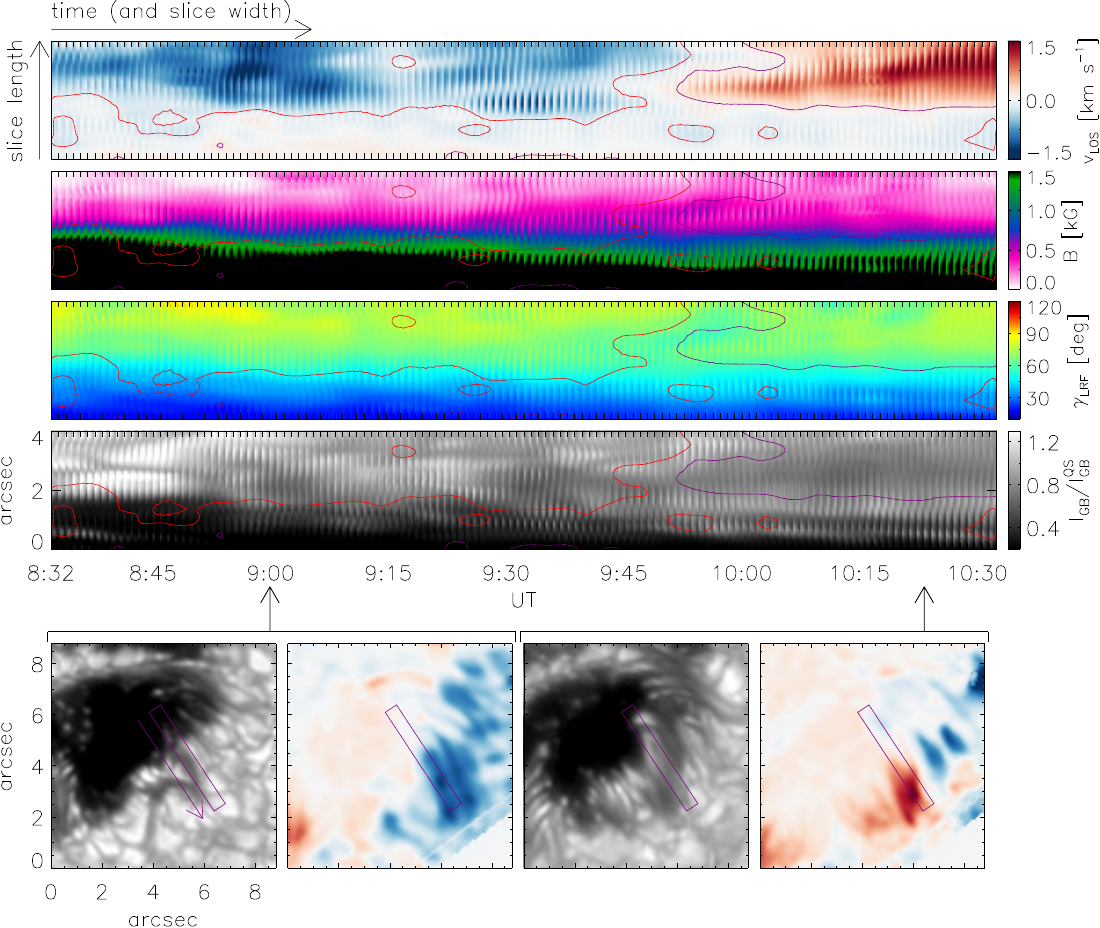}
\caption{Analogous to Fig.~\ref{timeslice_cs1} but showing the limb-side region where the granular pattern was observed before the penumbra formation. The bottom panels correspond to observations at 9:00\,UT (two left panels) and at 10:24\,UT (two right panels).}
\label{timeslice2}
\end{figure*} 

On a different segment of the limb-side boundary of the protospot, we analysed the granular pattern that precedes the formation of the Evershed flow. The evolution of this region is shown in Fig.~\ref{timeslice2}. As in the case of \ref{transient}, we were looking through the magnetic canopy of the protospot, and the physical parameters resulting from the inversion are influenced by this. 

At the displayed segment of the protospot boundary, in the continuum intensity maps we observed for the first 80 minutes granular patterns with granules smaller than those observed in the quiet-Sun regions. Despite being smaller, the mean intensity is comparable to the $I_\textrm{QS}$. During this time, we observed blueshifts. The structure of the LOS motions does not correspond to the granular pattern seen in the continuum intensity. In the dopplergram, the motions have a filamentary structure, and their lifetimes are longer than those of granular upflows. Such flows would correspond to the counter-Evershed flow described in Sect.~\ref{transient}. 

During the two studied hours, we did not detect any significant nor sudden changes in the configuration of the magnetic field along the slice in Fig.~\ref{timeslice2}. The gradients of the magnetic field strength and inclination along the slice length steadily decreased in time. During the first 80 minutes until 9:52\,UT, we did not detect any filaments with Evershed flows represented by redshifts. However, during the whole two hours, we observed short filaments on the protospot boundary that moved inwards and contained weak blueshifts. As explained in Sect.~\ref{transient}, these can be associated with upflows located in bright grains of such short filaments, where the outflow motion characterising the Evershed flow cannot be detected due to the viewing angle. 

The filaments carrying the Evershed flow started to be apparent in the intensity slices around 09:44\,UT as bright structures. We could not detect any LOS velocity due to the viewing angle. As the filaments lengthened in time, we started to detect redshifted LOS velocities around 09:52\,UT in parts of the filaments where the flow is horizontal and oriented towards the limb side. It is noteworthy that the Evershed flow sets in rapidly, within five to ten minutes. Until the end of the analysed period, the filaments lengthened and the Evershed flow became stronger.

\section{Discussion and conclusions}
\label{sec_conclusions}

We have identified three distinct types of magneto-convection-related motions prior to penumbra formation. These have been reported previously in other solar contexts unrelated to penumbra formation.

The bi-directional flows observed on the centre side and contained in one elongated granule, as seen in the intensity maps, resemble in many aspects the flux emergence event described in detail by \citet{Centeno2017}. In the case of the flux emergence described by \citet{Centeno2017}, the transversal magnetic field strength reached a maximum of 800~G. In our case, the horizontal field component is, at maximum, around 200~G. Higher field strength values can be found at the footpoints of the flow pattern, but the inner footpoint is influenced by the magnetic canopy of the protospot (see Fig.~\ref{fig:sketch_canopy}), and the outer footpoint is situated in the vicinity of another patch of positive polarity. In our case, the bright elongated granule is not as long and thin as in the case described by \citet{Centeno2017}. Based on this comparison and on the observations of orphan penumbra, where the horizontal field component is even higher \citep[above 1000~G;][]{Jurcak+etal2014} and where intensity structures are visually comparable to penumbral filaments, we surmise that the horizontal component of the magnetic field determines the width and length of the convective cell. 

Counter-Evershed flows have already been reported on the boundaries of the studied protospot by \citet{Schlichenmaier+etal2011}. There are other studies that have described observations of counter-Evershed flows in penumbra since then \citep{Louis2014, SiuTapia2017}. The recent statistical analysis of \cite{Castellanos2021} shows that counter-Evershed flows are common phenomena in the majority of active regions, although they are more frequently observed when the filaments carrying the counter-Evershed flow are related to sunspot light bridges \citep[see also][]{Kleint2013, Louis2020}. However, the typical lifetimes found by \cite{Castellanos2021} are significantly longer (most commonly around 7~hours) than the case studied in Sect.~\ref{transient}. Similar lifetimes in counter-Evershed flows have been found in sunspot MHD simulations (i.e. of a few hours) by \cite{2018_Siu-Tapia}. The big disparity between the counter-Evershed flow found in Sect.~\ref{transient} that lasts only approximately 30~minutes and the observed and simulated inflows that last a few hours suggests that the penumbral formation process may benefit from the existence of short-lived counter-Evershed flows. Based on a more detailed analysis of the counter-Evershed flows reported by \citet{Schlichenmaier+etal2011} at the centre side of the spot, we found that they actually correspond to the inner end of a bi-directional flow pattern.
                                                                       
It is well known that the magnetic boundaries of pores exceed those defined by intensity \citep{Keppens:1996} and that the size of the granules is influenced by the magnetic field strength \citep{Hirzberger2002, Lagg2014, Jurcak2017, 2017Vigeesh_B_effects}. Therefore, it is common to observe small granules near the boundaries of pores and protospots. Moreover, \cite{2017_Falco_granules_in_AR} showed that the effects of strong magnetic fields on inhibiting the convection are visible in the granules observed in light bridges and in umbral dots, where granular sizes are smaller than those reported in quiet-Sun regions and their mean continuum intensity is darker than in quiet-Sun granules. Yet granules spanning from small to quiet-Sun-like sizes are observed in plage regions with mean magnetic fields weaker than 700~G and quiet-Sun-like mean continuum intensities. As shown in \citet{Jurcak2017}, the intensity of granules is not influenced significantly by the magnetic field strength. We observed small but bright granules in regions with magnetic field strengths from 200~G to 700~G, depending on the distance from the protospot boundary, and with LRF inclinations of around 70$^\circ$ (the retrieved values are influenced by the LOS crossing the magnetic field canopy of the protospot; see Fig. \ref{fig:sketch_canopy}). Thus, the horizontal component is not strong enough (yet) to shape the granules into more elongated convective cells. However, LOS velocity maps of the same regions show a filamentary blueshifted structure that does not match a typical granular motion. The disagreement between the granular structure and the filamentary flow is possibly caused by the height difference between the formation regions of the continuum intensity and the \ion{Fe}{i} line. The lack of significant or sudden changes in the configuration of the magnetic field during the convective transformation from small granules to penumbral filament might be a combination of the effects of the LOS crossing the magnetic field canopy and the viewing angle.

Irrespective of the location where we observed the initial phases of the penumbra formation, the process always follows a similar pattern. We note that due to the viewing angle and LOS effects, the situation is best seen in Fig.~\ref{timeslice_cs1}, where the LOS is well aligned with the magnetic field on the umbral boundary, and we thus detected very weak flows along magnetic field lines. On the other hand, on the limb side, the LOS is close to being perpendicular to the magnetic field orientation, and the LOS velocities are thus hard to detect.

The process of penumbra formation starts within umbral areas with vertical fields below the critical threshold \citep{Jurcak_etal2015} and the appearance of short, bright filaments. \cite{Schmassmann+2021} has described the role of vertical fields in the inhibition of vigorous modes of magneto-convection in umbrae. These filaments carry weak flows oriented outwards, and the bright penumbral grains (heads) of these filaments move further into the umbra. 
As time progresses, the newly appearing filaments become longer and extend radially from the original location of the umbra--quiet-Sun boundary. Thus the sunspot expands to become larger in size. To some extent, the umbra-penumbra boundary also moves inwards to the umbra, as explained in \citet{Jurcak_etal2015}. 
As the penumbra forms, the radial gradient of the magnetic field strength and inclination decreases, eventually reaching values typically observed in stable sunspots (i.e. field strength and inclination around 2500~G and 30$^\circ$, respectively, on the inner penumbral boundary and around 600~G and 70$^\circ$ on the outer penumbral boundary). 

The initial appearance of the penumbral filaments can be explained by the MHD simulation of \citet{Rempel2011}, which describes how the mass loading along inclined field lines bends the magnetic field and creates filaments. The length of the filaments is given by the magnetic field inclination, so the more inclined the field initially is, the longer are the filaments that can be produced. However, it has not yet been observed if the actual process of filament formation is identical to this scenario. It is also possible that the short filaments are created by small-scale flux emergence in which the horizontal field is carried up to the photosphere by convective motions, as obtained with a numerical model of a forming sunspot by \cite{Chen_2017}. Yet, these authors found that most of the penumbral filaments form in the flux emergence region, while we find that the penumbral filaments mostly form in the other side of the spot. The main difference between these scenarios is that in the simulations, the penumbral filaments form from the pre-existing field, while in the second scenario, the penumbral filaments form due to additional flux emergence. Only in the case of the first scenario would we observe the re-emergence of the bent field lines (sea-serpent configuration). Given the complex configuration of the sunspot magnetic field, we cannot be sure if such a sea-serpent structure is indeed related to the re-emergence of the field lines that were bent by the mass loading or if it is just an emerging flux loop and part of the so-called background magnetic field that is observed around filaments everywhere in sunspot penumbra. We note that we are comparing the observations of forming filaments in a developing penumbra to the \citet{Rempel2011} MHD simulation of filaments in a stable penumbra.

 According to \citet{Rempel2011}, Evershed flows develop from the deflection of the (convective) upflow into a radial outflow by the action of Lorentz forces in a thin layer right below the  $\tau=1$ surface. We surmise that in the observations presented here, the bi-directional flows result from the elongation of (magneto-) convective cells by horizontal fields building up at the outskirts of the protospot. At this stage, the Lorentz force is not strong enough to drive an Evershed flow. 
During the following flux accumulation process \citep{Rezaei+etal2012}, the horizontal fields develop the necessary gradient (with height) to develop a Lorentz force that eventually deflects the emerging upflows into an Evershed flow. 

We find that the Evershed flow onsets in the umbra, and as it develops, it expands outwards into actual penumbral filaments. 
Physical conditions allowing for bi-directional flows and their evolution into Evershed flows will be investigated in future sunspot simulations.

\begin{acknowledgements}
 This work was supported by the Czech-German common grant, funded by the Czech Science Foundation under the project 23-07633K and by the Deutsche Forschungsgemeinschaft under the project BE 5771/3-1 (eBer-23-13412), and the institutional support ASU:67985815 of the Czech Academy of Sciences. The Vacuum Tower Telescope is operated by the Leibniz-Institut f\"ur Sonnenphysik (KIS) in Freiburg, at the Spanish Observatorio del Teide of the Instituto de Astrof\'{\i}sica de Canarias.
\end{acknowledgements}

\bibliographystyle{aa}
\bibliography{biblio}

\begin{thebibliography}{62}
\expandafter\ifx\csname natexlab\endcsname\relax\def\natexlab#1{#1}\fi

\bibitem[{{Bello Gonz{\'a}lez} {et~al.}(2019){Bello Gonz{\'a}lez}, {Jur{\v{c}}{\'a}k}, {Schlichenmaier}, \& {Rezaei}}]{BelloGlez+2019}
{Bello Gonz{\'a}lez}, N., {Jur{\v{c}}{\'a}k}, J., {Schlichenmaier}, R., \& {Rezaei}, R. 2019, in Astronomical Society of the Pacific Conference Series, Vol. 526, Solar Polariation Workshop 8, ed. L.~{Belluzzi}, R.~{Casini}, M.~{Romoli}, \& J.~{Trujillo Bueno}, 261

\bibitem[{{Bello Gonz{\'a}lez} {et~al.}(2012){Bello Gonz{\'a}lez}, {Kneer}, \& {Schlichenmaier}}]{BelloGlez+2012}
{Bello Gonz{\'a}lez}, N., {Kneer}, F., \& {Schlichenmaier}, R. 2012, \aap, 538, A62

\bibitem[{{Bello Gonz{\'a}lez} {et~al.}(2008){Bello Gonz{\'a}lez}, {Okunev}, \& {Kneer}}]{bello08}
{Bello Gonz{\'a}lez}, N., {Okunev}, O., \& {Kneer}, F. 2008, \aap, 490, L23

\bibitem[{{Bendlin} {et~al.}(1992){Bendlin}, {Volkmer}, \& {Kneer}}]{bendlin92}
{Bendlin}, C., {Volkmer}, R., \& {Kneer}, F. 1992, \aap, 257, 817

\bibitem[{{Borrero} {et~al.}(2007){Borrero}, {Bellot Rubio}, \& {M{\"u}ller}}]{borrero2007}
{Borrero}, J.~M., {Bellot Rubio}, L.~R., \& {M{\"u}ller}, D.~A.~N. 2007, \apjl, 666, L133

\bibitem[{{Borrero} {et~al.}(2014){Borrero}, {Lites}, {Lagg}, {Rezaei}, \& {Rempel}}]{borrero2014}
{Borrero}, J.~M., {Lites}, B.~W., {Lagg}, A., {Rezaei}, R., \& {Rempel}, M. 2014, \aap, 572, A54

\bibitem[{{Borrero} {et~al.}(2008){Borrero}, {Lites}, \& {Solanki}}]{Borrero:2008}
{Borrero}, J.~M., {Lites}, B.~W., \& {Solanki}, S.~K. 2008, \aap, 481, L13

\bibitem[{{Borrero} {et~al.}(2011){Borrero}, {Tomczyk}, {Kubo}, {Socas-Navarro}, {Schou}, {Couvidat}, \& {Bogart}}]{Borrero+etal2011}
{Borrero}, J.~M., {Tomczyk}, S., {Kubo}, M., {et~al.} 2011, \solphys, 273, 267

\bibitem[{{Castellanos Dur{\'a}n} {et~al.}(2021){Castellanos Dur{\'a}n}, {Lagg}, \& {Solanki}}]{Castellanos2021}
{Castellanos Dur{\'a}n}, J.~S., {Lagg}, A., \& {Solanki}, S.~K. 2021, \aap, 651, L1

\bibitem[{{Centeno} {et~al.}(2017){Centeno}, {Blanco Rodr{\'\i}guez}, {Del Toro Iniesta}, {Solanki}, {Barthol}, {Gandorfer}, {Gizon}, {Hirzberger}, {Riethm{\"u}ller}, {van Noort}, {Orozco Su{\'a}rez}, {Berkefeld}, {Schmidt}, {Mart{\'\i}nez Pillet}, \& {Kn{\"o}lker}}]{Centeno2017}
{Centeno}, R., {Blanco Rodr{\'\i}guez}, J., {Del Toro Iniesta}, J.~C., {et~al.} 2017, \apjs, 229, 3

\bibitem[{{Chen} {et~al.}(2017){Chen}, {Rempel}, \& {Fan}}]{Chen_2017}
{Chen}, F., {Rempel}, M., \& {Fan}, Y. 2017, \apj, 846, 149

\bibitem[{{Falco} {et~al.}(2017){Falco}, {Puglisi}, {Guglielmino}, {Romano}, {Ermolli}, \& {Zuccarello}}]{2017_Falco_granules_in_AR}
{Falco}, M., {Puglisi}, G., {Guglielmino}, S.~L., {et~al.} 2017, \aap, 605, A87

\bibitem[{{Gough} \& {Tayler}(1966)}]{Gough:1966}
{Gough}, D.~O. \& {Tayler}, R.~J. 1966, \mnras, 133, 85

\bibitem[{{Hirzberger} {et~al.}(2002){Hirzberger}, {Bonet}, {Sobotka}, {V{\'a}zquez}, \& {Hanslmeier}}]{Hirzberger2002}
{Hirzberger}, J., {Bonet}, J.~A., {Sobotka}, M., {V{\'a}zquez}, M., \& {Hanslmeier}, A. 2002, \aap, 383, 275

\bibitem[{{Hurlburt} \& {Rucklidge}(2000)}]{Hurlburt+Rucklidge2000}
{Hurlburt}, N.~E. \& {Rucklidge}, A.~M. 2000, \mnras, 314, 793

\bibitem[{{Jahn} \& {Schmidt}(1994)}]{Jahn+Schmidt1994}
{Jahn}, K. \& {Schmidt}, H.~U. 1994, \aap, 290, 295

\bibitem[{{Jur{\v c}{\'a}k} {et~al.}(2015){Jur{\v c}{\'a}k}, {Bello Gonz{\'a}lez}, {Schlichenmaier}, \& {Rezaei}}]{Jurcak+etal2015}
{Jur{\v c}{\'a}k}, J., {Bello Gonz{\'a}lez}, N., {Schlichenmaier}, R., \& {Rezaei}, R. 2015, \aap, 580, L1

\bibitem[{{Jur{\v c}{\'a}k} {et~al.}(2014){Jur{\v c}{\'a}k}, {Bellot Rubio}, \& {Sobotka}}]{Jurcak+etal2014}
{Jur{\v c}{\'a}k}, J., {Bellot Rubio}, L.~R., \& {Sobotka}, M. 2014, \aap, 564, A91

\bibitem[{{Jur{\v{c}}{\'a}k}(2011)}]{Jurcak2011}
{Jur{\v{c}}{\'a}k}, J. 2011, \aap, 531, A118

\bibitem[{{Jur{\v{c}}{\'a}k} {et~al.}(2014){Jur{\v{c}}{\'a}k}, {Bello Gonz{\'a}lez}, {Schlichenmaier}, \& {Rezaei}}]{Jurcak+etal2014b}
{Jur{\v{c}}{\'a}k}, J., {Bello Gonz{\'a}lez}, N., {Schlichenmaier}, R., \& {Rezaei}, R. 2014, \pasj, 66, S3

\bibitem[{{Jur{\v{c}}{\'a}k} {et~al.}(2015){Jur{\v{c}}{\'a}k}, {Bello Gonz{\'a}lez}, {Schlichenmaier}, \& {Rezaei}}]{Jurcak_etal2015}
{Jur{\v{c}}{\'a}k}, J., {Bello Gonz{\'a}lez}, N., {Schlichenmaier}, R., \& {Rezaei}, R. 2015, \aap, 580, L1

\bibitem[{{Jur{\v{c}}{\'a}k} {et~al.}(2017){Jur{\v{c}}{\'a}k}, {Lemmerer}, \& {van Noort}}]{Jurcak2017}
{Jur{\v{c}}{\'a}k}, J., {Lemmerer}, B., \& {van Noort}, M. 2017, in Fine Structure and Dynamics of the Solar Atmosphere, ed. S.~{Vargas Dom{\'\i}nguez}, A.~G. {Kosovichev}, P.~{Antolin}, \& L.~{Harra}, Vol. 327, 34--39

\bibitem[{{Jur{\v{c}}{\'a}k} {et~al.}(2018){Jur{\v{c}}{\'a}k}, {Rezaei}, {Gonz{\'a}lez}, {Schlichenmaier}, \& {Vomlel}}]{Jurcak_etal2018}
{Jur{\v{c}}{\'a}k}, J., {Rezaei}, R., {Gonz{\'a}lez}, N.~B., {Schlichenmaier}, R., \& {Vomlel}, J. 2018, \aap, 611, L4

\bibitem[{{Keppens} \& {Martinez Pillet}(1996)}]{Keppens:1996}
{Keppens}, R. \& {Martinez Pillet}, V. 1996, \aap, 316, 229

\bibitem[{{Kleint} \& {Sainz Dalda}(2013)}]{Kleint2013}
{Kleint}, L. \& {Sainz Dalda}, A. 2013, \apj, 770, 74

\bibitem[{{Lagg} {et~al.}(2014){Lagg}, {Solanki}, {van Noort}, \& {Danilovic}}]{Lagg2014}
{Lagg}, A., {Solanki}, S.~K., {van Noort}, M., \& {Danilovic}, S. 2014, \aap, 568, A60

\bibitem[{{Leka} {et~al.}(2009){Leka}, {Barnes}, {Crouch}, {Metcalf}, {Gary}, {Jing}, \& {Liu}}]{disambiguation_leka}
{Leka}, K.~D., {Barnes}, G., {Crouch}, A.~D., {et~al.} 2009, \solphys, 260, 83

\bibitem[{{Leka} \& {Skumanich}(1998)}]{Leka1998}
{Leka}, K.~D. \& {Skumanich}, A. 1998, \apj, 507, 454

\bibitem[{{Li} {et~al.}(2019){Li}, {Yan}, {Wang}, {Kong}, {Xue}, \& {Yang}}]{Li+2019}
{Li}, Q., {Yan}, X., {Wang}, J., {et~al.} 2019, \apj, 886, 149

\bibitem[{{Li} {et~al.}(2018){Li}, {Yan}, {Wang}, {Kong}, {Xue}, {Yang}, \& {Cao}}]{Li+2018}
{Li}, Q., {Yan}, X., {Wang}, J., {et~al.} 2018, \apj, 857, 21

\bibitem[{{Lim} {et~al.}(2013){Lim}, {Yurchyshyn}, {Goode}, \& {Cho}}]{Lim+etal2013}
{Lim}, E.-K., {Yurchyshyn}, V., {Goode}, P., \& {Cho}, K.-S. 2013, \apjl, 769, L18

\bibitem[{{Lindner} {et~al.}(2023){Lindner}, {Kuckein}, {Gonz{\'a}lez Manrique}, {Bello Gonz{\'a}lez}, {Kleint}, \& {Berkefeld}}]{Lindner_2023}
{Lindner}, P., {Kuckein}, C., {Gonz{\'a}lez Manrique}, S.~J., {et~al.} 2023, \aap, 673, A64

\bibitem[{{Lites} {et~al.}(1995){Lites}, {Low}, {Martinez Pillet}, {Seagraves}, {Skumanich}, {Frank}, {Shine}, \& {Tsuneta}}]{Lites_etal1995}
{Lites}, B.~W., {Low}, B.~C., {Martinez Pillet}, V., {et~al.} 1995, \apj, 446, 877

\bibitem[{{Louis} {et~al.}(2020){Louis}, {Beck}, \& {Choudhary}}]{Louis2020}
{Louis}, R.~E., {Beck}, C., \& {Choudhary}, D.~P. 2020, \apj, 905, 153

\bibitem[{{Louis} {et~al.}(2014){Louis}, {Beck}, {Mathew}, \& {Venkatakrishnan}}]{Louis2014}
{Louis}, R.~E., {Beck}, C., {Mathew}, S.~K., \& {Venkatakrishnan}, P. 2014, \aap, 570, A92

\bibitem[{{MacTaggart} {et~al.}(2016){MacTaggart}, {Guglielmino}, \& {Zuccarello}}]{MacTaggart+etal2016}
{MacTaggart}, D., {Guglielmino}, S.~L., \& {Zuccarello}, F. 2016, \apjl, 831, L4

\bibitem[{{Murabito} {et~al.}(2016){Murabito}, {Romano}, {Guglielmino}, {Zuccarello}, \& {Solanki}}]{Murabito+etal2016}
{Murabito}, M., {Romano}, P., {Guglielmino}, S.~L., {Zuccarello}, F., \& {Solanki}, S.~K. 2016, \apj, 825, 75

\bibitem[{{Panja} {et~al.}(2021){Panja}, {Cameron}, \& {Solanki}}]{Panja+2021}
{Panja}, M., {Cameron}, R.~H., \& {Solanki}, S.~K. 2021, \apj, 907, 102

\bibitem[{{Puschmann} {et~al.}(2006){Puschmann}, {Kneer}, {Seelemann}, \& {Wittmann}}]{puschmann06}
{Puschmann}, K.~G., {Kneer}, F., {Seelemann}, T., \& {Wittmann}, A.~D. 2006, \aap, 451, 1151

\bibitem[{{Rempel}(2011)}]{Rempel2011}
{Rempel}, M. 2011, \apj, 729, 5

\bibitem[{{Rempel}(2012)}]{Rempel2012}
{Rempel}, M. 2012, \apj, 750, 62

\bibitem[{{Rempel} {et~al.}(2009){Rempel}, {Sch{\"u}ssler}, {Cameron}, \& {Kn{\"o}lker}}]{Rempel2009}
{Rempel}, M., {Sch{\"u}ssler}, M., {Cameron}, R.~H., \& {Kn{\"o}lker}, M. 2009, Science, 325, 171

\bibitem[{{Rezaei} {et~al.}(2012){Rezaei}, {Bello Gonz{\'a}lez}, \& {Schlichenmaier}}]{Rezaei+etal2012}
{Rezaei}, R., {Bello Gonz{\'a}lez}, N., \& {Schlichenmaier}, R. 2012, \aap, 537, A19

\bibitem[{{Romano} {et~al.}(2013){Romano}, {Frasca}, {Guglielmino}, {Ermolli}, {Tritschler}, {Reardon}, \& {Zuccarello}}]{Romano+etal2013}
{Romano}, P., {Frasca}, D., {Guglielmino}, S.~L., {et~al.} 2013, \apjl, 771, L3

\bibitem[{{Romano} {et~al.}(2014){Romano}, {Guglielmino}, {Cristaldi}, {Ermolli}, {Falco}, \& {Zuccarello}}]{Romano+etal2014}
{Romano}, P., {Guglielmino}, S.~L., {Cristaldi}, A., {et~al.} 2014, \apj, 784, 10

\bibitem[{{Rucklidge} {et~al.}(1995){Rucklidge}, {Schmidt}, \& {Weiss}}]{Rucklidge1995}
{Rucklidge}, A.~M., {Schmidt}, H.~U., \& {Weiss}, N.~O. 1995, \mnras, 273, 491

\bibitem[{{Schlichenmaier} {et~al.}(2010{\natexlab{a}}){Schlichenmaier}, {Bello Gonz{\'a}lez}, {Rezaei}, \& {Waldmann}}]{Schlichenmaier+etal2010a}
{Schlichenmaier}, R., {Bello Gonz{\'a}lez}, N., {Rezaei}, R., \& {Waldmann}, T.~A. 2010{\natexlab{a}}, Astronomische Nachrichten, 331, 563

\bibitem[{{Schlichenmaier} {et~al.}(2011){Schlichenmaier}, {Gonz{\'a}lez}, \& {Rezaei}}]{Schlichenmaier+etal2011}
{Schlichenmaier}, R., {Gonz{\'a}lez}, N.~B., \& {Rezaei}, R. 2011, in IAU Symposium, Vol. 273, Physics of Sun and Star Spots, ed. D.~{Prasad Choudhary} \& K.~G. {Strassmeier}, 134--140

\bibitem[{{Schlichenmaier} {et~al.}(2010{\natexlab{b}}){Schlichenmaier}, {Rezaei}, {Bello Gonz{\'a}lez}, \& {Waldmann}}]{Schlichenmaier+etal2010b}
{Schlichenmaier}, R., {Rezaei}, R., {Bello Gonz{\'a}lez}, N., \& {Waldmann}, T.~A. 2010{\natexlab{b}}, \aap, 512, L1

\bibitem[{{Schlichenmaier} {et~al.}(2012){Schlichenmaier}, {Rezaei}, \& {Gonz{\'a}lez}}]{Schlichenmaier+etal2012}
{Schlichenmaier}, R., {Rezaei}, R., \& {Gonz{\'a}lez}, N.~B. 2012, in Astronomical Society of the Pacific Conference Series, Vol. 455, 4th Hinode Science Meeting: Unsolved Problems and Recent Insights, ed. L.~{Bellot Rubio}, F.~{Reale}, \& M.~{Carlsson}, 61

\bibitem[{{Schmassmann} {et~al.}(2021){Schmassmann}, {Rempel}, {Bello Gonz{\'a}lez}, {Schlichenmaier}, \& {Jur{\v{c}}{\'a}k}}]{Schmassmann+2021}
{Schmassmann}, M., {Rempel}, M., {Bello Gonz{\'a}lez}, N., {Schlichenmaier}, R., \& {Jur{\v{c}}{\'a}k}, J. 2021, \aap, 656, A92

\bibitem[{{Shimizu} {et~al.}(2012){Shimizu}, {Ichimoto}, \& {Suematsu}}]{Shimizu+etal2012}
{Shimizu}, T., {Ichimoto}, K., \& {Suematsu}, Y. 2012, \apjl, 747, L18

\bibitem[{{Simon} \& {Weiss}(1970)}]{Simon+Weiss1970}
{Simon}, G.~W. \& {Weiss}, N.~O. 1970, \solphys, 13, 85

\bibitem[{{Siu-Tapia} {et~al.}(2017){Siu-Tapia}, {Lagg}, {Solanki}, {van Noort}, \& {Jur{\v{c}}{\'a}k}}]{SiuTapia2017}
{Siu-Tapia}, A., {Lagg}, A., {Solanki}, S.~K., {van Noort}, M., \& {Jur{\v{c}}{\'a}k}, J. 2017, \aap, 607, A36

\bibitem[{{Siu-Tapia} {et~al.}(2018){Siu-Tapia}, {Rempel}, {Lagg}, \& {Solanki}}]{2018_Siu-Tapia}
{Siu-Tapia}, A.~L., {Rempel}, M., {Lagg}, A., \& {Solanki}, S.~K. 2018, \apj, 852, 66

\bibitem[{{Sobotka} {et~al.}(1999){Sobotka}, {Brandt}, \& {Simon}}]{Sobotka:1999b}
{Sobotka}, M., {Brandt}, P.~N., \& {Simon}, G.~W. 1999, \aap, 348, 621

\bibitem[{{Solanki} \& {Montavon}(1993)}]{Solanki:1993}
{Solanki}, S.~K. \& {Montavon}, C.~A.~P. 1993, \aap, 275, 283

\bibitem[{{Tiwari} {et~al.}(2013){Tiwari}, {van Noort}, {Lagg}, \& {Solanki}}]{Tiwari_etal2013}
{Tiwari}, S.~K., {van Noort}, M., {Lagg}, A., \& {Solanki}, S.~K. 2013, \aap, 557, A25

\bibitem[{{Vigeesh} {et~al.}(2017){Vigeesh}, {Jackiewicz}, \& {Steiner}}]{2017Vigeesh_B_effects}
{Vigeesh}, G., {Jackiewicz}, J., \& {Steiner}, O. 2017, \apj, 835, 148

\bibitem[{{Wentzel}(1992)}]{Wentzel1992}
{Wentzel}, D.~G. 1992, \apj, 388, 211

\bibitem[{{Zuccarello} {et~al.}(2014){Zuccarello}, {Guglielmino}, \& {Romano}}]{Zuccarello+etal2014}
{Zuccarello}, F., {Guglielmino}, S.~L., \& {Romano}, P. 2014, \apj, 787, 57

\bibitem[{{Zwaan}(1987)}]{Zwaan1987}
{Zwaan}, C. 1987, \araa, 25, 83

\end{thebibliography}

\end{document}